\documentclass[preprint,12pt]{elsarticle}

\usepackage{graphicx}
\usepackage{graphics}
\usepackage{rotating}
\usepackage{longtable}
\usepackage{lscape}
\usepackage{epsfig}
\usepackage{rotating}
\usepackage{ifthen}
\usepackage{multicol}

\usepackage{amssymb}
\usepackage{xspace}

\usepackage{lineno}

\newcommand{\ife}[3]{\ifthenelse{\equal{#1}{#2}}{#3}{{}}}
\newcommand{\otwo}{$\mathrm{O_2}$}
\newcommand{\ntwo}{$\mathrm{N_2}$}
\newcommand{\chfour}{$\mathrm{CH_4}$}

\begin{document}
\linenumbers

\begin{frontmatter}

\title{Study of a zirconium getter for purification of xenon gas}

\author{A.~Dobi}
\author{D.~S.~Leonard\fnref{DL}}
\author{C.~Hall}
\author{L.~Kaufman}
\author{T.~Langford}
\author{S.~Slutsky}
\author{Y.-R.~Yen}
\address{Department of Physics, University of Maryland, College Park MD, 20742 USA}

\fntext[DL]{Now at University of Seoul, Seoul, South Korea}

\begin{abstract}

Oxygen, nitrogen and methane purification efficiencies for a common zirconium 
getter are 
measured in 1050 Torr of xenon gas. Starting with impurity 
concentrations near 
$10^{-6}$ g/g, the outlet impurity level 
is found to be less 
than $120 \cdot 10^{-12}$ g/g for $\mathrm{O_2}$ and 
less than $950 \cdot10^{-12}$ g/g  for $\mathrm{N_2}$.
For methane we find residual contamination of the purified gas
at concentrations varying over three orders of magnitude,
depending on the purifier temperature and the gas flow rate.
A slight reduction 
in the purifier's methane efficiency is observed after 13 mg of this 
impurity has been absorbed, which we attribute to partial exhaustion
of the purifier's capacity for this species. 
We also find that the purifier's ability to absorb $\mathrm{N_2}$
and methane can be extinguished long before any decrease in $\mathrm{O_2}$ 
performance is observed, and slower flow rates should be 
employed for xenon purification due to the cooling effect
that the heavy gas has on the getter.


\end{abstract}


\end{frontmatter}


\section{Introduction}
\label{sec:Intro}

Liquefied noble gases have been widely adopted as a particle 
detection medium in recent years. 
These materials are attractive
candidates for detectors
due to their low ionization potential, high 
scintillation efficiency, and, in the case of xenon, high density 
and stopping power \cite{AprileXeRev09}.
The successful operation of these detectors requires that the noble 
liquid should be almost entirely free of non-inert
impurities such as $\mathrm{O_2}$ and $\mathrm{H_2O}$ 
because these species impair the transport of both
scintillation light and ionization charge.

Achieving extremely 
low levels of electronegative impurities is of particular concern
for ionization detectors. For example, the presence of $\mathrm{O_2}$ at
only the part-per-billion level will give rise to 
a free electron lifetime of 
several hundred microseconds, which is comparable to the
drift time of many present-day detectors \cite{Bakale}. Next generation 
experiments will require drift distances an order of 
magnitude larger than those currently existing, and therefore a 
proportional increase in the purity must be obtained. Reliably 
achieving such purities is a significant 
technical barrier that these experiments must overcome.

While a variety of noble gas purification technologies have been 
shown 
to give good results \cite{Pokachalov,Benetti,Bettini,Ichige,Buckley,Carugno,Bolotnikov,Aprile91},
the most common technology in use today 
is the heated zirconium getter. Zirconium is an effective absorbent
because its surface bonds with virtually any 
non-noble gas species \cite{Stojilovic}, including 
$\mathrm{O_2}$, $\mathrm{H_2O}$, $\mathrm{N_2}$, $\mathrm{CO_2}$, 
and \chfour. 
These getters are operated at a temperature
of several hundred degrees Celsius, which encourages 
impurities that are bonded to surface sites to diffuse into 
the bulk, leaving the 
surface available for additional gettering. The 
efficiency improves significantly as the temperature 
increases due to the decrease in the 
diffusion time \cite{Giorgi,Baker93,Watanabe98}.

Zirconium getters designed for noble gas purification are commercially 
available, and the SAES Monotorr series in particular has been widely 
adopted for particle detection applications. SAES specifies and 
guarantees the performance of 
the Monotorr for helium and argon service, but they have not
performed any measurements in xenon due to the high cost of this 
rare gas. We are not aware of any previous measurements of the
performance of the Monotorr in gaseous xenon 
with sensitivity at the part-per-billion level.
 
It is reasonable to expect that the getter 
performance in xenon will be similar to 
that of argon and helium, and xenon detectors have seen good 
results while using these purifiers. 
Nevertheless, it is conceivable
that the getter temperature may be negatively affected
by the high mass of this noble gas, and this could 
reduce the purifier's efficiency.
 
Here we report measurements of the xenon purification efficiency 
of the SAES Monotorr PS4-MT3 \cite{SAES}. 
We have studied the performance of this zirconium getter 
for removal of 
$\mathrm{O_2}$  and $\mathrm{N_2}$, two common 
electronegative impurity species found
in small quantities in virtually all sources of xenon. 
We have also studied the removal of methane, 
a less common contaminant which is present in the atmosphere at 
only the part-per-million level. 
Methane is not electronegative, so it does
not present a problem for charge transport. 
In fact, a few percent of methane is sometimes added to xenon 
in order to increase the charge drift speed \cite{kirill-drift-velocity}.  
Methane can also have negative consequences, however, 
because it quenches the dimer molecules responsible
for the fluorescence emission \cite{kirill-scintillation}.

We have chosen to study methane removal from xenon for 
several
reasons. First, methane is largely inert, so it represents one 
of the most challenging cases for a xenon 
purifier \cite{Watanabe98,Kawano,Klein98,Antoniazzi}. Second, 
our measurement technique is sensitive to methane at concentrations
as low as $60 \cdot 10^{-12}$ g/g, 
and therefore we can measure
the purifier performance very accurately for this gas. Finally, 
methane isomers such as $\mathrm{^{14}CH_4}$ or $\mathrm{CH_3T}$
are candidates for internal beta calibration
sources for large liquid xenon experiments. Since these sources
would need to be removed from the xenon at the conclusion of
the calibration run, it is important to 
understand quantitatively the performance and limitations 
of the zirconium getter for this impurity.

\section{Measurement technique}

\begin{figure}[t!]\centering
\includegraphics[width = 120 mm]{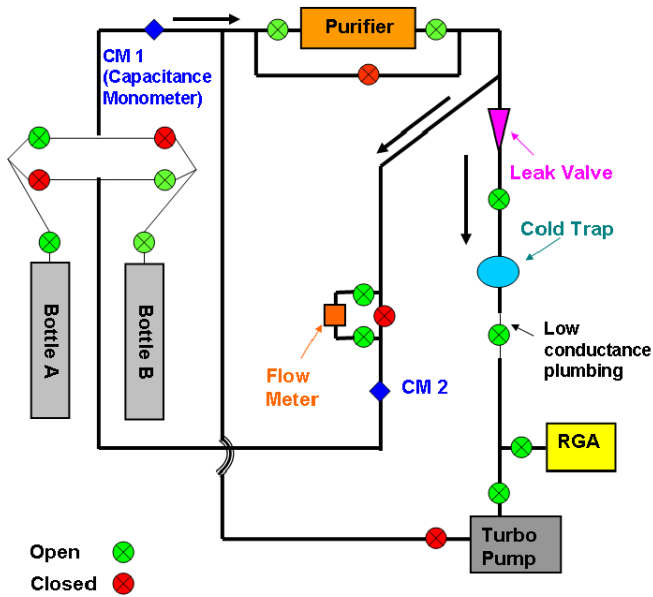}
\caption{The xenon handling system for the experiment. 
The impurity concentration of the gas is analyzed 
downstream of the purifier using the coldtrap and RGA.}
\label{fig:system}
\end{figure}

The xenon handling system for our experiments is shown in Fig. 
\ref{fig:system}.
Xenon gas is stored in one of two 15.73 liter aluminum storage 
cylinders and 
supplied to the system through a regulator. The gas is cryopumped
into the second cylinder by immersing it in liquid nitrogen.
The plumbing is designed to allow the gas to flow either through 
the purifier (purify mode) or around the purifier (bypass mode). 
Flow rates were measured using a mass flow meter from MKS. The absolute 
scale of the flow meter is calibrated by transferring a known mass of 
xenon gas at a constant rate through the meter.

To measure the purity of the xenon gas before and after purification, 
we use an RGA/coldtrap technique described in Ref.  \cite{coldtrap}. 
We use a leak valve to sample the gas composition at the 
output of the purifier. The chemical composition of the sampled gas 
is measured by a residual gas analyzer (RGA, model: SRS RGA200), a mass spectroscopy 
device which operates at low pressures ($10^{-5}$ Torr or less). 
To improve the sensitivity of the method, we remove most of the 
xenon from the low pressure gas sample  
with a liquid nitrogen coldtrap before it reaches the RGA. 
The impurities which we study in 
this paper ($\mathrm{N_2}$, $\mathrm{O_2}$, and $\mathrm{CH_4}$) 
have relatively high vapor 
pressures compared to xenon at 77 ~K, and they survive the cold 
trap in large enough quantities to be measured. 
The selective rejection of the xenon 
gas allows the sampling rate of the leak valve to be vastly 
increased without saturating the RGA with a high partial pressure 
of the bulk xenon gas. The measurement is calibrated by preparing 
xenon gas samples which are spiked with known amounts of the various 
impurities under study. Using this method the RGA's impurity 
sensitivity is boosted from 
$\sim 10^{-5}$ 
to $60 \cdot 10^{-12}$ g/g for methane. 
For $\mathrm{N_2}$ the limit of detection is less than 
$10^{-9}$ g/g and for $\mathrm{O_2}$ it is about $120 \cdot 10^{-12}$ g/g. 
The limitations for the different species are due to 
their specific background levels.
All signals used for data analysis were background subtracted to compensate for outgassing in the plumbing from the coldtrap to the RGA. We expect the boost in the RGA's performance to be an order of magnitude greater if the plumbing from the coldtrap to the RGA is baked.
We note here that the concentration 
of impurities in the xenon gas is observed to be linear with the 
partial pressure measured by the RGA, so that a factor of ten 
reduction in the partial pressure corresponds to a factor of 
ten reduction in the concentration. See Ref.  \cite{coldtrap} for 
additional details.

We use 3.25 kg of xenon initially purified 
to the part-per-billion level using a 
Monotorr purifier. After this initial purification, the 
used zirconium cartridge was replaced with a new one. 
Helium with
a purity level of 99.999\% is added to the xenon
with a concentration
of $8 \cdot 10^{-9}$ g/g  to serve as a tracer gas during 
measurements. The tracer provides a useful calibration 
signal because it is unaffected by both the purifier and the cold
trap. The purification efficiencies were calculated 
from the initial and final values of the impurity-to-helium ratio. 
We normalize our measurements to the helium
signal in order to make small corrections for changes in 
gas pressure and flow rate. Note that these corrections are 
typically at the level of only a few percent. To prepare spiked samples impurities of interest are injected into a known volume connected to a pressure gauge. Using purified xenon the impurity is flushed out of the injection volume and is cryopumped into a storage bottle to allow for full mixing. Uncertainties in the prepared mixtures are within 5\% and are determined from the uncertainty in the injection volume and the error in the pressure gauge. 

For each test the xenon handling system is first filled with 
roughly 1050 Torr of static xenon gas spiked with $10^{-6}$ g/g 
of the specific impurity under study. For this initial fill, the 
purifier is 
operated in bypass mode, so that the leak valve 
samples the unpurified gas. The leak valve to the coldtrap 
is opened and the RGA partial-pressure measurements are allowed to 
stabilize. This calibrates the partial-pressure measurements 
in terms of the known concentration of the impurity in the spiked
gas sample.
Then the plumbing is switched to purify mode and xenon is made to flow
from its supply bottle to a collection bottle passing  
through the purifier at a fixed flow rate.  The 
output pressure of the purifier is maintained at 1050 Torr $\pm$ 
3\% during data acquisition to maintain a constant leak rate into the 
coldtrap and RGA. The partial pressure of the impurity is allowed
several minutes to stabilize, and then its value is recorded.
At the conclusion of the measurement, the input to 
the cold-trap is closed to get a background measurement. In some 
tests, using lower leak rates, the leak-valve is opened further at 
the end of the measurement to increase its sensitivity before 
closing the input. 

For highest sensitivity measurements, it is crucial to minimize the 
backgrounds in the coldtrap. This can be accomplished by 
making the unpurified gas measurement after the purified gas 
measurement, because the unpurified gas can
contaminate the coldtrap and RGA plumbing with residual background 
levels for some period of time. For these tests the background
level is measured first while the coldtrap remains 
closed and the xenon handling system is filled with purified gas. 
Then the purified xenon is allowed to flow through the purifier 
for 20 minutes at a fixed rate before the leak 
valve to the cold trap is opened. If no change in the impurity 
level is seen, 
we place a limit on the purified signal by assuming that
the residual impurity concentration in the purified gas is less 
than 20\% of the observed background level. Finally we 
normalize the measurement (or limit) to the known impurity 
concentration in the spiked xenon gas by bypassing the purifier and 
recording the resulting impurity level with the RGA.

Only one data set is taken per day, and the purifier is left at
its operational temperature overnight. This is done as a precaution 
to make sure that 
the absorbed impurities have time to diffuse into the bulk of the 
getter, an issue that we discovered while using an older, mostly
exhausted purifier. See section \ref{sec:lifetime} for more
details.

\section{Purifier efficiency results}

We define the purifier efficiency to be the fraction of a given 
impurity which is removed by the purifier in a single pass 
under specified flow and temperature conditions. The inefficiency
is the fraction of the impurity which remains after a single 
pass.

\subsection{Methane purification efficiency}
 
Among the impurities that we measure, the purifier has the smallest 
total absorption capacity for methane, so we tested methane purification 
first. During the course of these experiments, the purifier was
exposed to roughly 16 mg 
of methane, which is less than the 10\% of the cartridge's estimated capacity. 
To create the xenon-methane mixture we use a methane supply bottle 
which has a stated purity level of 99.999\%.
In the following, we report results for two purifier temperatures, 
400 $^\circ$C and 450 $^\circ$C. Unfortunately, we do not have physical access to the 
zirconium in the getter, so we cannot measure these 
temperatures directly. Instead, we quote these temperatures based upon 
the manufacturer's specifications \cite{SAESprivate}. Please contact SAES technical support before modifying the getter temperature.

\begin{figure}[t!]\centering
\includegraphics[width = 120 mm]{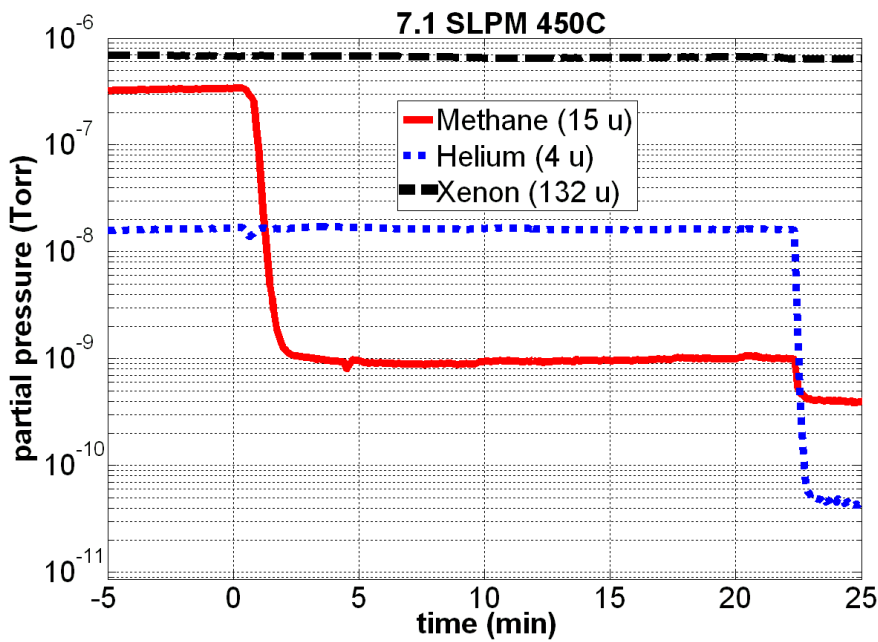}
\caption{A plot of methane purification data at 450 $^\circ$C and 7.1 SLPM.
The leak valve rate is set to 43 Torr$\cdot$L/min. See text for further
details.}
\label{fig:7SLPM450}
\end{figure}

A typical dataset is shown in Figure \ref{fig:7SLPM450}. For this
data the xenon was spiked with $1  \cdot  10^{-6}$ g/g of methane, 
and the purifier temperature is 450 $^\circ$C. 

This particular experiment proceeded as follows.
Prior to the data shown in the plot, the coldtrap was cooled
to liquid nitrogen temperature, and the leak valve was opened.
The xenon plumbing system is initially filled in bypass mode
with the
unpurified xenon gas, and the RGA detects a partial pressure 
of about $3  \cdot  10^{-7}$ Torr of methane. At time t = 0, 
the plumbing is switched to purify mode, and the xenon gas is 
made to flow
at 7.1 SLPM. The helium is unaffected by the purifier, so its 
partial pressure remains constant, but the 
methane partial pressure is seen to drop by about a factor
of 300 over the next few minutes. The xenon partial 
pressure remains constant due to the action of the coldtrap.
At 22 minutes, the leak valve is closed, and 
the helium and methane levels drop to their background 
values, while the xenon pressure again remains constant.

\begin{figure}[t!]\centering
\includegraphics[width = 120 mm]{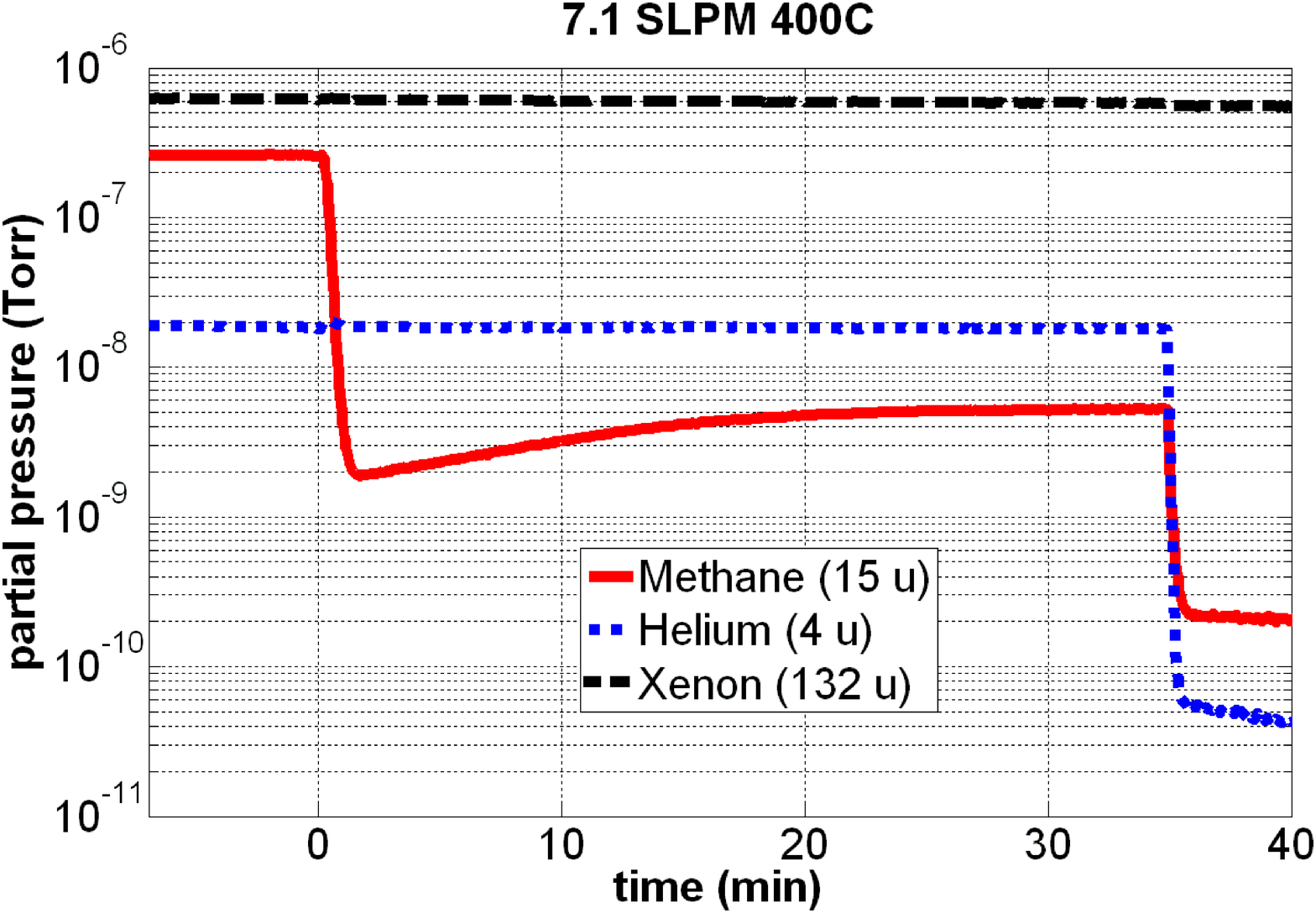}
\caption{Methane purification data at 400 $^\circ$C and 
a flow rate of 7.1 SLPM. The leak valve rate is 43 
Torr$\cdot$L/min.
We attribute the slow rise in the methane
signal in this data 
to the cooling effect of the xenon on the getter. 
Compare to the 450 $^\circ$C data shown in Fig. \ref{fig:7SLPM450}, 
where no rise is evident. Note that 400 $^\circ$C is the 
default getter temperature.}
\label{fig:7SLPM400}
\end{figure}

Figure \ref{fig:7SLPM400} shows a similar dataset, taken
at 7.1 SLPM and 400 $^\circ$C.  
In the 400 $^\circ$C dataset, we see the methane
drop sharply as expected when switched to purify mode, and
then rise slowly by a factor of $\sim$2.5 over the next
20 minutes. In the 450 $^\circ$C dataset (shown in Fig. \ref{fig:7SLPM450}) 
the slow rise is not present.
We infer from these results that the getter efficiency is reduced 
at 400 $^\circ$C by the 
cooling effect of the flowing xenon gas, while at 450 $^\circ$C the 
getter is able to maintain its full efficiency.

\begin{figure}[t!]\centering
\includegraphics[width = 120 mm]{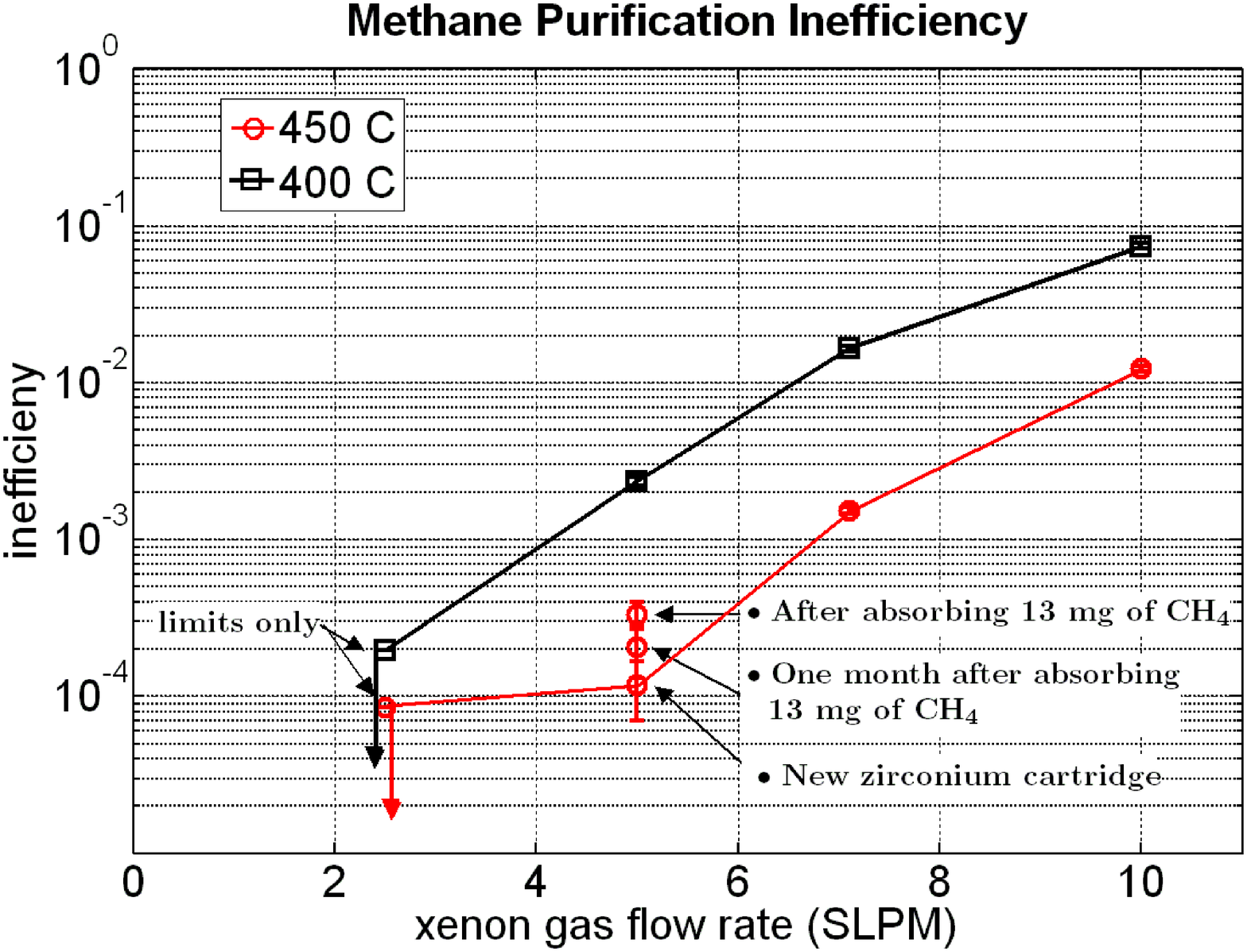}
\caption{Methane purification inefficiency as a function 
of flow rate, for two purifier temperatures. The inefficiency
is defined as the fraction of the incoming 
methane which remains at the purifier output. }
\label{fig:methane_inefficiency}
\end{figure}

\begin{table}[t!]
\begin{center}
\begin{tabular}{|c|c|c|c|c|c|c|}
\hline
 &    Gas flow rate  & Temp. & Init. conc.& Final conc. & 1-pass eff.& 1-pass ineff.\\
Species &   [SLPM] & [C] & $10^{-6}$ [g/g] & $10^{-9}$ [g/g] & [\%] & [\%] \\
\hline  
\chfour & 2.5 $\pm .05$  & 400 & 1.2 & $<0.24$        & $>99.98$           & $<0.02$  \\
\chfour & 5.0 $\pm .05$  & 400 & 1.8 & $4.3 \pm 0.36$ & $99.76 \pm 0.02$   & $0.24 \pm 0.02$   \\ 
\chfour & 7.1 $\pm .05$ & 400 & 1.0 & $17 \pm 0.07$  & $98.35 \pm 0.07$   & $1.65 \pm 0.07$   \\
\chfour & 10 $\pm .05$  & 400 & 1.2 & $89.5 \pm 3.6$ & $92.53 \pm 0.30$   & $7.47 \pm 0.30$  \\
\hline
\chfour & 2.5 $\pm .05$ & 450 & 1.0 & $<0.085$           & $>99.991$           & $<0.009$   \\
\chfour & 5.0 $\pm .05$  & 450 & 1.8 & $0.162 \pm 0.09$   & $99.988 \pm 0.005$  & $0.012 \pm 0.005$ \\
\chfour & 7.1 $\pm .05$  & 450 & 1.0 & $0.190 \pm 0.08$   & $99.846 \pm 0.008$  & $0.15  \pm 0.008$  \\
\chfour & 10 $\pm .05$  & 450 & 1.8 & $22.1  \pm 1.1$    & $98.77  \pm 0.06$   & $1.23  \pm 0.06$   \\
\hline
\otwo  & 12.6 $\pm .05$ & 400 & 1.0 & $<0.120$ & $>99.988$ & $<0.012$ \\
\hline
\ntwo  & 12.6 $\pm .05$ & 400 & 1.0 & $<0.954$ & $>99.905$ & $<0.095$ \\
\hline    
\end{tabular}
\end{center}
\vspace{0.25in}
\caption{Purification results for $\mathrm{O_2}$ , $\mathrm{N_2}$, and methane.
Initial concentrations are accurate to within 5\%. The error in inefficiency is calculated from the final and initial background-subtracted signals, each normalized to helium. Each data point was determined by averaging a stabilized signal over a period of one minute. The typical standard deviation in the signal was found to be 2\%.}

\label{tab:data}
\end{table}

Figure \ref{fig:methane_inefficiency} shows how the purification inefficiency
for methane depends on gas flow rate and purifier temperature. These 
results are also listed in Table~\ref{tab:data}.
As expected, the purifier performed best for methane
at the higher getter 
temperature and at lower flow-rates. At 450 $^\circ$C and 2.5 SLPM 
the methane purification
efficiency is better than 99.99\% and consistent with background, 
while at 400 $^\circ$C and 10 SLPM the efficiency drops to 92.5\%.

\subsection{$\mathrm{N_2}$ and $\mathrm{O_2}$ purification efficiency}

The getter cartridge was also tested with $10^{-6}$ g/g of $\mathrm{O_2}$ 
and $10^{-6}$ g/g of $\mathrm{N_2}$ at 400 $^\circ$C and at a flow rate of 
12.6 SLPM (the maximum achievable flow rate in our system). 
In purify mode, both $\mathrm{O_2}$ and $\mathrm{N_2}$ dropped to levels consistent 
with background, leading to upper limits for the one-pass purification 
efficiency of $> 99.988$\% for $\mathrm{O_2}$ and 
$> 99.905 $\% for $\mathrm{N_2}$. These efficiencies correspond 
to output purity levels of less than $120 \cdot 10^{-12}$ g/g for $\mathrm{O_2}$
and less than $950 \cdot 10^{-12}$ for $\mathrm{N_2}$. 
We infer from the methane purification data 
that lower flow rates and higher temperatures 
will achieve even better purification efficiencies. 
For $\mathrm{N_2}$ and $\mathrm{O_2}$ the purification 
results in xenon gas are consistent with those quoted by the 
manufacturer for argon and helium.

\section{Purifier Lifetime and Saturation}
\label{sec:lifetime}

Other studies have observed 
eventual decrease 
in purification efficiencies due to capacity 
depletion \cite{Baker93}.
To track the performance of the SAES PS4-MT3 as its capacity 
begins to be depleted, we collected the first and last methane 
datasets under identical conditions at a flow rate of 
5.0 SLPM and a getter temperature of 450 $^\circ$C.  
We find that after the zirconium has absorbed 13 mg of methane,
the getter inefficiency increases from 0.01\% to 0.03\% (see 
Figure \ref{fig:methane_inefficiency}). 
Furthermore, 
the measurement at 5.0 SLPM was repeated a third time after leaving 
the purifier with 1000 Torr of xenon at its operational temperature for one month without use. 
The purifier's performance appears to improve somewhat after this 
interim period. However, the
final methane signal for this dataset was only a few percent above 
the background, and it is difficult rule out a systematic 
effect of this magnitude over such a long time period.

\subsection{Purifier life-status light}

The SAES PS4-MT3-R purifier is equipped with a life-status light which 
indicates when the absorption capacity of the getter has been 
exhausted. The control electronics for the purifier infers 
the status of the getter by measuring its resistance. As a 
practical matter, it is important to note that this measurement 
is sensitive primarily 
to $\mathrm{O_2}$ and other oxygen containing impurities, since zirconium oxide has a higher 
resistance than pure zirconium. Other zirconium compounds, such 
as zirconium nitride, do not cause a dramatic change in the 
resistance of the getter, and their presence will not be 
detected by the resistance measurement. Therefore the purifier's 
ability to remove $\mathrm{N_2}$ and methane may be depleted long before 
the life-status light changes from 'good'.

\begin{figure}[t!]\centering
\includegraphics[width = 90 mm]{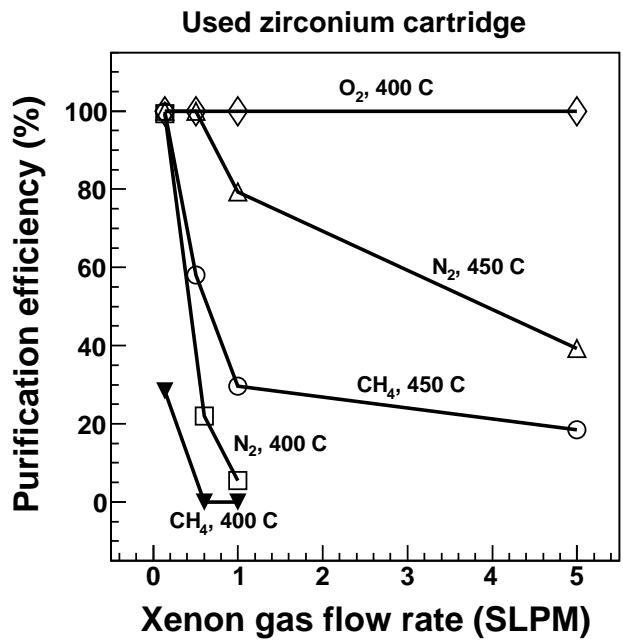}
\caption{Purification results with an old zirconium cartridge.
The purifier remains effective for $\mathrm{O_2}$ removal, but $\mathrm{N_2}$
and methane performance is seriously diminished. The purifier's
life-status light still indicates 'good'. }
\label{fig:old-zirconium-cartridge}
\end{figure}

In fact, when we initially tested our purifier with an older 
getter cartridge that had been in use in our lab for about 
two years,  we observed that the purifier's performance with 
respect to $\mathrm{N_2}$ and methane was dramatically diminished 
although the life-status was indicated as 'good'.
At 0.5 SLPM the used 
getter cartridge was still effectively removing $\mathrm{O_2}$ while 
allowing nearly 100\% of $\mathrm{N_2}$ and methane to pass through
at 400 C, as shown in Fig. \ref{fig:old-zirconium-cartridge}. 
Note that this purifier model, when new, 
is rated for use up to a maximum
flow rate of 20 SLPM for argon and helium.  
We also note that this model 
is designed to remove about 40 g of $\mathrm{O_2}$, 2-3 g 
of $\mathrm{N_2}$ and about 200 mg of methane in its 
lifetime \cite{SAESprivate}.

Before replacing the used getter cartridge, we performed a final 
test comparing its purification efficiencies at 400 $^\circ$C to 450 C, 
for $\mathrm{N_2}$ and methane. 
The xenon used in the test had $4 \cdot 10^{-6}$ g/g  
of methane and about $30 \cdot 10^{-6}$ g/g of $\mathrm{N_2}$. 
The results are shown in Figure \ref{fig:old-zirconium-cartridge}.
These tests confirmed that the used 
cartridge performs better at the higher getter temperature
and at lower flow rates. 
We also observed that the performance of the 
used getter cartridge would improve somewhat after leaving 
the getter at its operational temperature for one or more 
days. This observation is consistent with the expected 
decrease in the impurity diffusion rate as the bulk zirconium
becomes saturated near the end of its life. 

In total, our results illustrate the utility of having a 
dedicated gas purity measurement device to monitor the 
performance of the getter.

\section{Summary}

$\mathrm{O_2}$ and $\mathrm{N_2}$ are removed from xenon 
very efficiently by the getter.
Concentrations at the outlet are less than $120 \cdot 10^{-12}$ g/g and 
$955 \cdot 10^{-12}$ g/g, respectively, and are consistent with background. 
The corresponding single-pass purification efficiencies are greater
than 99.99\% and 99.9\% respectively. These values were measured
at a flow rate of 12.6 SLPM, 
and at the default getter temperature of 400 $^\circ$C. We 
expect that performance at lower flow rates and higher temperatures 
is even better. These results are consistent with the manufacturer's
specifications for the getter in argon and helium.

Methane removal from xenon is most effective at lower flow rates
and higher getter temperatures. The removal efficiency is better 
than 99.99\% at 450 $^\circ$C and 2.5 SLPM and drops to 92.5\% at 400 $^\circ$C 
and 10 SLPM. We attribute the decrease in getter effectiveness at 
higher flow rates to the cooling effect xenon gas has on 
the getter. 

For flow rates of five SLPM or higher, the residual methane 
concentration at the outlet of the getter can be decreased 
by about a factor of ten by increasing the temperature 
from 400 $^\circ$C to 450 $^\circ$C.

After absorbing 13 mg of methane, we see residual methane
in the purified gas increase by a factor of three.
We attribute this effect to a partial saturation of the purifier's methane
absorption sites.

The getter can become saturated with $\mathrm{N_2}$ and methane 
while remaining effective for $\mathrm{O_2}$ removal. For xenon use, the manufacturer's 
life-status light is not well suited for determining when  
$\mathrm{N_2}$ or methane saturation has occurred.
Therefore, additional techniques should be employed to determine
the status of the getter with respect to these species.
In many
systems of interest for particle detection, we expect 
the $\mathrm{N_2}$
capacity of the purifier to be exhausted before the
$\mathrm{O_2}$ capacity. 

Since the 
getter performance in xenon improves at higher temperatures, 
it seems reasonable that 
higher temperatures should always be employed when
purifying this heavy noble gas. Alternatively, a 
xenon gas pre-heater could be introduced immediately upstream of 
the purifier to reduce or eliminate the cooling effect. 
We have not attempted this strategy in our lab, but we would expect
good results based on our experience with this purifier.

\section{Acknowledgments}
We thank Marco Succi of SAES Getters for many helpful discussions.
This work was supported by National Science Foundation award numbers
0810495 and 0919261.

%

\bibliographystyle{elsarticle-num}
\bibliography{purifier}

\end{document}